\documentclass[final,5p,times,twocolumn,sort&compress]{elsarticle}

\usepackage{graphicx}
\usepackage{array}
\usepackage{caption}
\usepackage{subcaption}
\usepackage{float}
\usepackage{subfig}

\usepackage{amsmath}
\usepackage{slashed}

\newcommand{\be}{\begin{equation}}
\newcommand{\ee}{\end{equation}}

\newcommand{\p}{\vec{p}}
\newcommand{\pp}{\vec{p}^{\, \prime}\!}

\newcommand{\ord}{\mathcal{O}}

\newcommand{\cfunij}{G_{ij}(\p ,t;\, \Gamma)}
\newcommand{\cfunijZ}{G_{ij}(0 ,t;\, \Gamma_4)}
\newcommand{\cfunijmu}{G_{ij}(\pp ,\p;\, t_2, t_1;\, \Gamma')}
\newcommand{\cfunijmuZZ}{G_{ij}(0 ,0;\, t_2, t_1;\, \Gamma_3)}
\newcommand{\cfunA}{G^{\alpha}(\p ,t;\Gamma)}
\newcommand{\cfunAmu}{G^{\alpha}(\pp ,\p;\, t_2, t_1;\, \Gamma')}
\newcommand{\RA}{R^{\alpha}(\pp ,\p; \, \Gamma',\Gamma)}

\newcommand{\chii}{\chi_{i}}
\newcommand{\chiDj}{\bar{\chi}_{j}}

\newcommand{\cfunxij}{\langle \, \Omega \, | \, \chii(x) \, \chiDj(0) \, | \, \Omega \, \rangle}
\newcommand{\cfunxijmu}{\langle \Omega \, | \, \chii(x_2) \, \mathcal{O}(x_1) \, \chiDj(0) \, | \, \Omega \rangle}

\newcommand{\rCoupp}{\langle \alpha, \, p, s \, | \, \chiDj(0) \, | \, \Omega \rangle}
\newcommand{\lCouppp}{\langle \Omega | \, \chii(0) \, | \, \beta, \, p', s' \rangle}

\newcommand{\vertexAB}{\langle \, \beta, p', s' \, | \, \mathcal{O}(0) \, | \, \alpha, p,s  \rangle}

\journal{Physics Letters B}

\begin{document}

\begin{frontmatter}



\title{Variational Approach to the Calculation of $g_A$}


\author[CSSM]{Benjamin~J.~Owen\corref{Ben}}
\ead{benjamin.owen@adelaide.edu.au}
\cortext[Ben]{Corresponding author}
\author[CSSM]{Jack~Dragos}
\author[CSSM]{Waseem~Kamleh}
\author[CSSM]{Derek~B.~Leinweber}
\author[CSSM]{M.~Selim~Mahbub}
\author[CSSM]{Benjamin~J.~Menadue}
\author[CSSM]{James~M.~Zanotti}

\address[CSSM]{Special Research Centre for the Subatomic Structure of Matter,\\
School of Chemistry \& Physics, University of Adelaide, South Australia, 5005, Australia}

\begin{abstract}
A long standing problem in Lattice QCD has been the discrepancy between the experimental and calculated values for the axial charge of the nucleon, $g_A \equiv G_{A}(Q^2=0)$. Though finite volume effects have been shown to be large, it has also been suggested that excited state effects may also play a significant role in suppressing the value of $g_A$. In this work, we apply a variational method to generate operators that couple predominantly to the ground state, thus systematically removing excited state contamination from the extraction of $g_A$. The utility and success of this approach is manifest in the early onset of ground state saturation and the early onset of a clear plateau in the correlation function ratio proportional to $g_A$. Through a comparison with results obtained via traditional methods, we show how excited state effects can suppress $g_A$ by as much as 8\% if sources are not properly tuned or source-sink separation are insufficiently large.
\end{abstract}


\end{frontmatter}


\section{Introduction}
In recent years, lattice calculations have taken a tremendous step towards simulating QCD at the physical point. Algorithmic and technological developments have allowed simulations to probe at or near physical quark masses on increasingly larger volumes, with finer lattice spacings and vastly increased statistics. Calculations of the ground state spectrum have yielded results consistent to within a few percent of their physical values with well controlled systematic errors \cite{Fodor:2012,Hoelbling:2011}. Naturally the next step has been to strive for this level of precision for the matrix elements of these states. Despite the remarkable consistency between lattice and experimental data for the pion form factor $F_{\pi}(Q^2)$, a complete description of other hadronic states, particularly the nucleon, has proven to be remarkably challenging \cite{Hagler:2009,Alexandrou:2010:HS}.

The most notable shortfall is for the nucleon axial charge, $g_{A} \equiv G_{A}(Q^2=0)$. In principle $g_A$ should be relatively simple to calculate. Being an iso-vector quantity, disconnected loop contributions are absent and as we have direct access to $G_A(0)$, we circumvent the need for extrapolations in $Q^2$. Unfortunately, the lattice values for $g_A$ to date have been consistently lower than the experimental value by as much as 10--15\% \cite{Lin:2012}. In an effort to account for these discrepancies, several studies have carefully examined the systematic errors present in the calculation \cite{Sasaki:2003,Edwards:2005,Khan:2006,Hagler:2007,Lin:2008,Yamazaki:2008,Lin:2007,Bratt:2010,Alexandrou:2010:Ax,Pleiter:2011,Dinter:2011,Hall:2012,Capitani:2012:Ax,Horsley:2013}. In this letter we will focus on the role of excited state effects.

Recently there has been an increased effort to understand and reduce the impact of excited states on form factor calculations. In computing these quantities, it is well understood that to ensure excited state contributions to the correlation function are sufficiently suppressed, one needs large Euclidean time separations between operators. To choose a suitable time separation one should identify the time slices where the correlation functions take on their asymptotic form. For the two commonly used sequential source techniques, this is a relatively simple procedure for the fixed current method. One simply chooses a current insertion time, $t_C$, once the asymptotic behaviour is observed in the two-point correlator. Results are extracted from the data once the asymptotic behaviour is observed in the three-point correlator.

For the fixed sink method, one requires knowledge of the asymptotic behaviour of the three-point correlator \textit{a priori}. Unfortunately, the temptation to use earlier sink times in order to obtain more precise results is inescapable. These results can suffer from excited state contaminations, even if a plateau is observed with $t_C$. In refs. \citep{Lin:2008,Dinter:2011}, it was found that for certain matrix elements, eg. $\langle x \rangle$, the source-sink separations often used in the literature were not sufficiently large to suppress excited state effects. Nonetheless, as we move ever closer to the physical point one is naturally forced to choose earlier sink times as the signal degrades much quicker.

To counter this issue, new techniques are being devised to try and control the sub-leading terms to the three point correlator. The use of the summation method \cite{Capitani:2012:Ax, Capitani:2012:ES} has shown improvement upon the conventional approach, but the underlying excited states contributions are still present. It is not hard to imagine situations where these still impact significantly and alter the final result.

In this paper we take a somewhat different approach. Rather than reduce the impact of excited states through Euclidean time evolution, we seek to separate them out from the ground state at the source and sink. Drawing upon the techniques developed for excited state spectroscopy calculations, we will use the variational approach to construct interpolating fields that couple with individual energy eigenstates and use these to isolate the desired matrix elements \cite{Owen:2012,Menadue:2012}. An analogous approach has been presented in \cite{Bulava:2011,Bernardoni:2012} for the study of $B^* \rightarrow B \pi$ transitions and in \cite{Maurer:2012} for the study of the axial charges of Nucleon excited states. Here we apply it specifically to $g_A$ to remove excited state contributions.

This letter is organized as follows. In Section 2 we will examine the variational method in the context of excited state spectroscopy and then outline how this method can be applied to the calculation of hadronic matrix elements. Section 3 outlines the details of this calculation. In Section 4 we present our results and compare our variational method with the traditional, single-operator approach to the calculation of $g_A$. Section 5 is a cost-benefit discussion for the variational method. Finally we provide our concluding remarks in Section 6.

\section{Variational Method for Matrix Elements}
The `variational method' \cite{Michael:1985,Luscher:1990} is a well established approach for determining the excited state hadron spectrum. It is based on the creation of a matrix of correlation functions in which different superpositions of excited state contributions are linearly combined to isolate the energy eigenstates. A diversity of excited state superpositions is central to the success of this method.

Starting from a basis of operators $\left\lbrace \, \chi_{i}(x) \, |\,i=1,\ldots,N\,\right\rbrace$, we construct a correlation matrix of two-point correlation functions,
\be
G_{ij}(\p;\, t;\, \Gamma) = \sum_{\vec{x}} e^{-i \p\cdot\vec{x}} \: \textrm{tr} \, \Big( \, \Gamma \, \cfunxij \, \Big) \: .
\ee

Due to the discrete nature of the lattice, we can decompose these correlation functions into a discrete sum over energy eigenstates,
\be
\label{eq:twoptSD}
\cfunij = \sum_{\alpha} e^{-E_{\alpha}(\p)\,t}\: Z^{\alpha}_{i}(\p)\, \bar{Z}^{\alpha}_{j}(\p) \, \textrm{tr} \, \left( \frac{\Gamma(\slashed{p}+m_{\alpha})}{2E_{\alpha}(\p)} \right) \: ,
\ee
where the parameters $Z^{\alpha}_{i}(\p)$ are the coupling strengths of the interpolators $\chi_i(x)$ with the energy eigenstate of mass $m_{\alpha}$ and $\Gamma$ projects out the desired parity. We choose new operators $\phi^{\alpha}(x)$ to be linear combinations
\be
\phi^{\alpha}(x) = \sum_{i} v_{i}^{\alpha} \, \chi_{i}(x) \: , \:\:\: \bar{\phi}^{\alpha}(x) = \sum_{j} u_{j}^{\alpha} \, \bar{\chi}_{j}(x) \: ,
\ee
with a suitable choice of coefficients $v^{\alpha}_{i}$ and $u^{\alpha}_{j}$, such that these interpolators couple to a single energy eigenstate,
\be
\label{eq:orthog}
\langle \, \Omega \, | \, \phi^\beta(0) \, | \, \alpha, p, s \, \rangle = \delta^{\alpha\beta} \, \mathcal{Z}_{\alpha}(\p) \, \sqrt{\frac{m_{\alpha}}{E_{\alpha}(\p)}} \, u(p,s) \: .
\ee
From Eqs. \eqref{eq:twoptSD} and \eqref{eq:orthog} we find that the necessary values for $v^{\alpha}_{i}$ and $u^{\alpha}_{j}$ are the solutions of the following eigenvalue equations
\be
\label{eq:leftEV}
v^{\alpha}_{i}(\p) \, [G(\p,t_0+\Delta t) \, (G(\p,t_0))^{-1}]_{ij} = c^{\alpha} \, v^{\alpha}_{j}(\p) \: ,
\ee
\be
\label{eq:rightEV}
[ (G(\p,t_0))^{-1} \, G(\p,t_0+\Delta t)]_{ij} \, u^{\alpha}_{j}(\p) = c^{\alpha} \, u^{\alpha}_{i}(\p) \: ,
\ee
where the eigenvalue $c^{\alpha} = e^{-m_\alpha \Delta t}$.

It is important to note that both \eqref{eq:leftEV} and \eqref{eq:rightEV} are evaluated for a given momentum $\vec{p}$ and so the diagonalisation condition is only satisfied when we project with the relevant coefficients as follows:
\be
v^{\alpha}_{i}(\p) \, \cfunij \, u^{\beta}_{j}(\p) \propto \delta^{\alpha\beta} \: .
\ee
Thus the two-point correlation function for the state $|\,\alpha, p \rangle$ is
\be
\cfunA \equiv v^{\alpha}_{i}(\p) \, \cfunij \, u^{\alpha}_{j}(\p) \: .
\ee
We can extract the mass $m_\alpha$ from $G^{\alpha}(\p=0,t)$ in the standard way.

To understand how we can utilise the variational method for use in form factor calculations, we must first identify the terms present in the three-point correlation function,
\begin{multline}
G_{ij}(\pp ,\p;\, t_2, t_1;\, \Gamma') = \sum_{\vec{x}_1,\vec{x}_2} e^{-i \pp\cdot\vec{x}_2} \, e^{+i (\pp-\p)\cdot\vec{x}_1} \\
\textrm{tr} \, \Big( \, \Gamma' \, \cfunxijmu \, \Big) \: .
\end{multline}
where $\mathcal{O}(x)$ is the current operator to be inserted.
Sandwiching the current between two complete sets of states we end up with three terms, the vertex amplitude, $\vertexAB$, and the coupling terms $\lCouppp$ and $\rCoupp$,
\begin{multline}
G_{ij}(\pp ,\p;\, t_2, t_1;\, \Gamma') = \sum_{\alpha, \, \beta} \, e^{-E_{\beta}(\pp)(t_2-t_1)} \, e^{-E_{\alpha}(\p) t_1} \\
Z^{\beta}_{i}(\pp) \, \bar{Z}^{\alpha}_{j}(\p) \sqrt{\frac{m_{\alpha}\,m_{\beta}}{E_{\alpha}(\p)\,E_{\beta}(\pp)}} \, \textrm{tr} \, \bigg( \, \Gamma' \, \sum_{s',s} u(p',s') \\
\langle \, \beta, p', s' \, | \, \mathcal{O}(0) \, | \, \alpha, p, s \, \rangle \, \bar{u}(p,s) \, \bigg) \: .
\end{multline}
The coupling parameters take the same form as they did in the calculation of the two-point correlator with two key differences. The inclusion of a current means that the initial and final momenta need not be the same. Furthermore, there also exists the possibility that the initial and final energy eigenstates are not the same. That is, the current can induce a transition between states. For this calculation the necessary expression is
\be
\cfunAmu = v^{\alpha}_{i}(\pp) \, \cfunijmu \, u^{\alpha}_{j}(\p) \: .
\ee
To isolate the matrix element from the three-point function, we construct a ratio in the standard way. In this work we shall use the ratio defined in \cite{Hedditch:2007}. For the state $\alpha$ the necessary ratio is,
\be
\label{eq:ratio}
\RA = \sqrt{\frac{G^{\alpha}(\pp ,\p;\, t_2, t_1;\, \Gamma')\,G^{\alpha}(\p ,\pp;\, t_2, t_1;\, \Gamma')}{G^{\alpha}(\p ,t_2;\, \Gamma)\,G^{\alpha}(\pp ,t_2;\, \Gamma)}} \: .
\ee

Key to this approach is the utilisation of a basis of operators in which there is diversity in the overlap with various excited states. As there are a limited number of local bilinear operators for a given $J^{PC}$, a great deal of work has been made by various groups in increasing the number of available operators. Here we choose to use fermion source and sink smearing as a method of extending our operator basis, as outlined in \cite{Mahbub:2009,Mahbub:2010:Qu}.

\section{Calculation Details}
For this calculation we make use of the PACS-CS (2+1)-flavour dynamical-QCD gauge field configurations \cite{Aoki:2008} made available through the ILDG \cite{Beckett:2009}. These configurations are generated using a non-perturbatively $\ord(a)$-improved Wilson fermion action and Iwasaki gauge action. The value $\beta = 1.90$ results in a lattice spacing $a = 0.091\textrm{ fm}$, determined via the static quark potential. With dimensions $32^3$ $\times$ $64$, these ensembles correspond to a spatial length of $L = 2.9 \textrm{ fm}$. As the intention of this paper is to examine whether the variational approach is an improvement upon traditional techniques, we will consider only the light quark mass that corresponds to $m_\pi \approx 290\textrm{ MeV}$. The resulting value of $m_{\pi}L = 4.26$ is comparable to the values used by most groups.

In this work we are primarily interested in isolating the ground state and so have chosen to use a small variational basis upon which to perform our correlation matrix analysis. We use gauge-invariant Gaussian smearing in the spatial dimensions only with smearing fraction $\alpha=0.7$ \cite{Mahbub:2010:R}. We consider four levels of smearing with the optimal choice found in \cite{Mahbub:2010:R}, these being 16, 35, 100 and 200, applied to the standard, local proton interpolator
\[
\chi_1(x) = \epsilon^{abc} [ u^{a}{}^{T}(x) C \gamma_5 d^{b}(x) ] \, u^{c}(x) \, ,
\]
thus allowing for construction of a correlation matrix of dimension up to 4 $\times$ 4. In table 1 we list the rms-radii for our choice of smearing parameters. We choose to use variational parameters $t_0 = 18$ and $\Delta t = 2$, again taken from \cite{Mahbub:2010:R}, where it was found that this choice produced best balance between systematic and statistical uncertainties.

\begin{table}[h!]
	\centering
	\caption{The rms radii for the various levels of smearing considered in this work.}
	\begin{tabular}{cc}
	\hline
	\noalign{\smallskip}
	Sweeps of smearing & rms radius (fm) \\
	\hline
	\noalign{\smallskip}
	16  & 0.216 \\
	35  & 0.319 \\
	100 & 0.539 \\
	200 & 0.778 \\
	\hline
	\end{tabular}
\end{table}

To extract the nucleon axial charge we are interested in the matrix element $\langle \, p(p', s') \, | \, A_{\mu}^{ud} \, | \, n(p, s) \, \rangle$ where $A_{\mu}^{ud} = \bar{u} \gamma_{\mu} \gamma_5 d$. This vertex can be expressed via two independent form factors, the axial form factor $G_A(Q^2)$ and the induced pseudoscalar form factor $G_P(Q^2)$, as
\begin{multline}
\langle \, p(p', s') \, | \, A_{\mu}^{ud} \, | \, n(p, s) \, \rangle = \left( \frac{m^2}{E_{p'}E_p} \right)^{1/2} \\ \bar{u}_p(p',s') \left[\gamma_{\mu}\gamma_5 G_{A}(Q^2) + \, \gamma_5 \frac{q_{\mu}}{2m} G_{P}(Q^2) \right] u_n(p,s) \: ,
\end{multline}
where $q_\mu = p_{\mu}' - p_{\mu}$ and $Q^2 = - \, q^2$. Using isospin symmetry, one can show that for the flavour-changing current $A_{\mu}^{ud}$, the matrix element is equivalent to that of the iso-vector current $A_{\mu}^{u-d}$,
\[
\langle \, p(p', s) \, | \, A_{\mu}^{ud} \, | \, n(p, s) \, \rangle = \langle \, p(p', s) \, | \, A_{\mu}^{u-d}  \, | \, p(p, s) \, \rangle \: ,
\]
where $A_{\mu}^{u-d} = \bar{u} \gamma_{\mu} \gamma_5 u -\bar{d} \gamma_{\mu} \gamma_5 d$. We choose to calculate $g_A$ using $\mathcal{O} = A_{\mu}^{u-d}$.

As we are interested in $G_A(Q^2=0)$, it suffices to consider the case where the incoming and outgoing momenta are the same, in particular we choose to work in the nucleon rest frame as this will provide the smallest statistical uncertainties. This will mean that the left and right eigenvectors required to project out the three-point function will now correspond to the same momenta.

We choose to insert our fermion source at $t_0 = 16$. For the calculation of the three-point functions we use a local axial vector current calculated using a sequential source technique with the current held fixed and inserted at $t_C = 21$, at the onset of asymptotic behaviour for the projected two-point function. We choose to use $\mu = 3$ for the current with the corresponding projection matrix being $\Gamma' = \Gamma_3 = \Gamma_4 \, \gamma_5 \, \gamma_3$, where $\Gamma_4 \equiv \frac{1}{2}(I + \gamma_0)$. The value for the axial renormalization constant $Z_A = 0.781(20)$ was determined non-perturbatively in \citep{Aoki:2010} using a Schr\"{o}dinger functional scheme.

The resulting expression from which we extract $g_A$ is the ratio of the eigenstate-projected three-point and two-point functions,
\be
\label{gAratio}
g_A^{CM} = \frac{v^{0}_{i}(0) \, \cfunijmuZZ \, u^{0}_{j}(0)}{v^{0}_{i}(0) \, \cfunijZ \, u^{0}_{j}(0)} \: .
\ee
As a comparison, we also evaluate $g_A$ using a single correlator from smeared source to point sink and smeared source to smeared sink. These are indicative of results one would extract from a traditional approach.

\section{Results}

\begin{figure}
	\centering
	\begin{subfigure}[b]{\columnwidth}
		\centering
		\includegraphics[width=\textwidth]{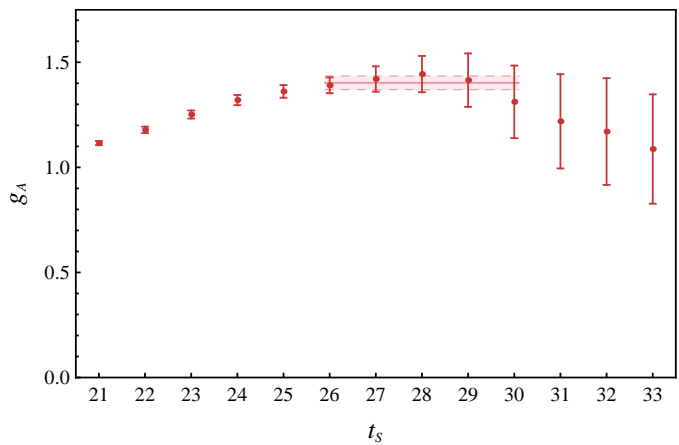}
		\caption{Smeared source to point sink}
		\label{pt}
	\end{subfigure} \\
	\vspace{3mm}
	\begin{subfigure}[b]{\columnwidth}
		\centering
		\includegraphics[width=\textwidth]{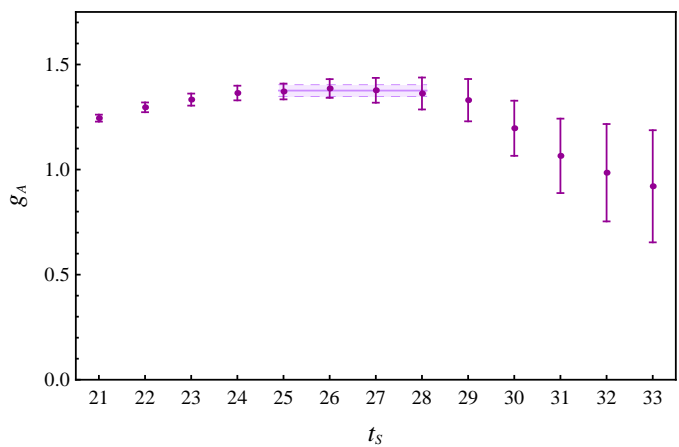}
		\caption{Smeared source to smeared sink}
		\label{sm}
	\end{subfigure} \\
	\vspace{3mm}
	\begin{subfigure}[b]{\columnwidth}
		\centering
		\includegraphics[width=\textwidth]{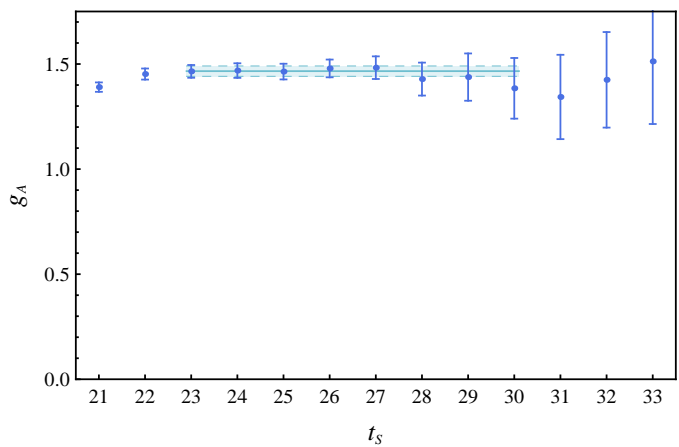}
		\caption{Correlation matrix approach}
		\label{CM}
	\end{subfigure}
	\caption{A comparison of un-renormalized $g_A$ as a function of sink time. The first two figures are using the traditional approach of \textit{smeared} source $\rightarrow$ \textit{point} sink and, \textit{smeared} source $\rightarrow$ \textit{smeared} sink, both for 35 sweeps of smearing. The final figure is the result from a $4 \times 4$ correlation matrix.}
\end{figure}

In Fig.~1 we present the bare values of $g_A$ with increasing sink time $t_s$ following the current insertion at $t_C = 21$ for the smeared source to point sink, smeared source to smeared sink (both with 35 sweeps of smearing) and our variational method respectively. Between the traditional approach (upper two plots) and the variational approach (bottom plot), we can see significant differences in the overall shape of the correlation function ratio.

For the smeared source to point sink (upper plot) the Euclidean time suppression of excited state contributions manifests itself as a steady increase in the extracted value. This trend in the data does not have a clear endpoint and so there is no definite plateau. Guided by the $\chi^2_{\textrm{dof}}$ obtained via a covariance matrix analysis, the earliest time slice one could consider is $t_S = 25$, but what is clear is that we are forced to consider fit windows uncomfortably close to regions dominated by noise.

By smearing the sink as well as the source, there is a definite improvement in the quality of the plateau. The excited state behaviour is again present as a steady increase in the value of $g_A$, but somewhat suppressed. In this case there is a definite plateau observed at $t_S = 24$, which is supported by the $\chi^2_{\textrm{dof}}$. Unfortunately, this is again somewhat close to the region where signal is lost to noise.

In Fig.~1 (c) we see quite a different situation. Our variational approach yields extremely clean results with rapid ground state dominance. The systematic rise in the data is no longer present and the onset of the plateau is within two time slices of the current insertion.

\begin{figure}[t]
	\centering
	\includegraphics{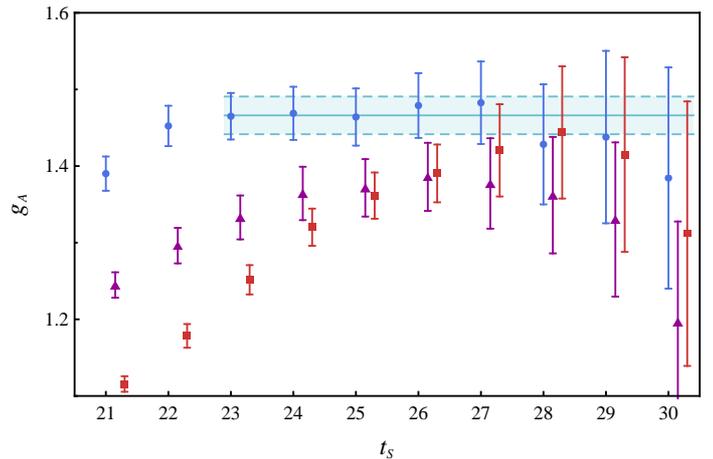}
	\caption{An overlay of the results from fig.~1. The data sets have been offset from the time slice for clarity -- the circles (blue) are the results for the variational approach, the triangles (purple) are the \textit{smeared} source $\rightarrow$ \textit{smeared} sink, while the squares (red) are the \textit{smeared} source $\rightarrow$ \textit{point} sink. The fitted value from the variational approach has been included (blue shaded region) to highlight where the traditional approach is consistent with the improved method.}
\end{figure}

In Fig.~2 we have overlaid the three datasets to highlight the excited state behaviour between the traditional and variational approach. If we look carefully at the variational approach, we can see that some excited state contamination is present immediately after the current, but this is short lived. It is worth noting that this is a consequence of the limited size of our variational basis. As is highlighted in \citep{Bernardoni:2012}, an $n \times n$ correlation matrix allows one to isolate out the $n$ lightest states in the given channel and so the sub-leading contributions will come from the $n^{th}+1$ state. In the case of the ground state, these contributions will be short-lived due to the large mass splitting between the ground state and $n^{th}+1$ excited state. If one were to construct a basis whose dimension was the number of states in the given channel, then it would be possible to completely isolate each state.

\begin{figure}[t]
	\centering
	\includegraphics{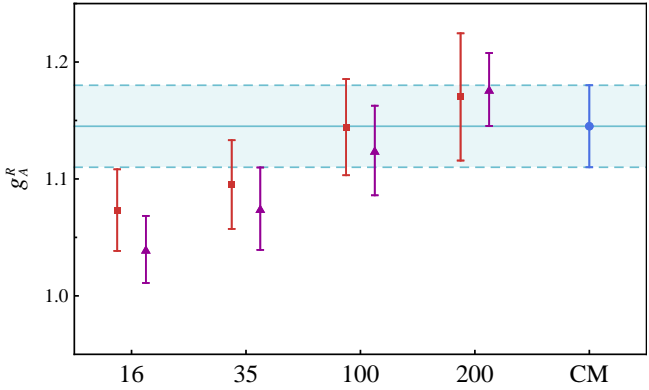}
	\caption{Comparison of the renormalized value of $g_A$. The first four pairs of points are the results for the conventional, point sink (squares) and smeared sink (triangles) approach with increasing levels of smearing to the right. The rightmost point (circle) is the result extracted using variational approach. There is a clear dependence on the level of smearing to the extracted result.}
\end{figure}

What is of most concern in Fig.~2 is the lack of overlap between the results of the traditional approach and those of our variational method at $t_s = 24$ and 25 where good fits can be made. In Table 2 we list those fits, for the three data sets with the strict criterion that the $\chi^2_{\textrm{dof}}$ lies between 0.800 and 1.200. In both data sets employing the traditional approach, we can obtain good fits with small uncertainties if we choose to begin fitting around $t_S=25$ or $26$, but find that the results are significantly small. As we move the fit window to later times, the central value increases. Between 25--30 and 28--30 we observe a systematic variation of 6\% in the value $g_A$. A consistent result can be extracted from these datasets if we choose to fit at later time slices around $t_S = 28$, but the resulting values have unattractively large uncertainties, as they are close to the onset of noise. It is clear that in this case, we have little control over the excited state systematics. In contrast, for the various fit windows on the data from the variational approach we find the variation in the fits is considerably smaller than the smallest statistical uncertainty.

\begin{table*}[t]
	\centering
	\caption{Un-renormalized values of $g_A$ from fit windows which give a covariance matrix based $\chi^2_{\textrm{dof}}$ between 0.800 and 1.200. The datasets are identified as (a) Traditional approach with point sink, (b) Traditional approach with smeared sink and (c) Variational approach, where for the traditional approach we have selected the 35 sweeps of smearing. We note how the value of $g_A$ increases for the traditional approach as we move the fit window to later times. In contrast, the variational approach is stable across all windows with the desired $\chi^2_{\textrm{dof}}$.}
	\begin{tabular}{ccccccccccc}
		\multicolumn{3}{c}{(a)} && \multicolumn{3}{c}{(b)} && \multicolumn{3}{c}{(c)} \\
	\cline{1-3} \cline{5-7} \cline{9-11}
	\noalign{\smallskip}
	Fit Window & $g_A$ & $\chi^2_{\textrm{dof}}$ && Fit Window & $g_A$ & $\chi^2_{\textrm{dof}}$ && Fit Window & $g_A$ & $\chi^2_{\textrm{dof}}$ \\
	\cline{1-3} \cline{5-7} \cline{9-11}
	\noalign{\smallskip}
	25 -- 27 & 1.38(3) & 1.168  && 24 -- 30 & 1.36(3) & 1.161 && 23 -- 30 & 1.47(3) & 0.848 \\
	\noalign{\smallskip}
	25 -- 30 & 1.38(4) & 1.100  && 24 -- 31 & 1.36(3) & 1.104 && 23 -- 31 & 1.47(3) & 0.818 \\
\cline{5-7} \cline{9-11}\noalign{\smallskip}
	25 -- 31 & 1.38(3) & 0.951  && 25 -- 28 & 1.38(3) & 0.926 && 24 -- 29 & 1.47(2) & 0.848 \\
\cline{1-3}\noalign{\smallskip}
	26 -- 27 & 1.40(3) & 0.808  && 25 -- 29 & 1.37(3) & 0.812 && 24 -- 30 & 1.47(2) & 0.988 \\
\cline{5-7}\noalign{\smallskip}
	26 -- 30 & 1.40(3) & 1.077  && 26 -- 30 & 1.37(4) & 1.100 && 24 -- 31 & 1.47(4) & 0.932 \\
\cline{9-11}\noalign{\smallskip}
	26 -- 31 & 1.40(4) & 0.902  && 26 -- 31 & 1.36(4) & 0.952 && 25 -- 29 & 1.47(3) & 0.951 \\
\cline{1-3} \cline{5-7}\noalign{\smallskip}
	27 -- 31 & 1.41(4) & 1.011  && 27 -- 31 & 1.33(4) & 1.148 && 25 -- 30 & 1.47(2) & 1.120 \\
\cline{1-3} \cline{5-7}\noalign{\smallskip}
	28 -- 30 & 1.42(6) & 1.129  && 28 -- 31 & 1.30(10) & 1.082 && 25 -- 31 & 1.47(2) & 1.040 \\
\cline{1-3} \cline{5-7} \cline{9-11}\noalign{\smallskip}
	29 -- 31 & 1.35(7)& 0.994  &&          &         &       && 26 -- 28 & 1.47(2) & 1.091 \\
\cline{1-3}\noalign{\smallskip}	
	        &         &         &&          &         &       && 26 -- 29 & 1.47(2) & 1.184 \\      		\noalign{\smallskip}
   	        &         &         &&          &         &       && 26 -- 31 & 1.47(2) & 1.146 \\
\cline{9-11}
	\end{tabular}
\end{table*}

It is worth considering how the level of smearing affects the extracted value of $g_A$. In Fig.~3, we present the renormalized $g_A$ considering each of the four smearings used to construct our variational basis. What we find is a dependence on the level of smearing used in the calculation. It appears that for low levels of smearing the extracted result can be significantly lower, with the smallest level of smearing differing by up to 8\% from our improved, variational result. From this evidence, it is clear that if the smearing level is not properly tuned at the source and sink, then excited state effects significantly impact the extracted result for $g_A$.

In principle, one could tune the smearing so that the optimal overlap is observed with the ground state. By using a point source propagator and tuning the smearing through the sink via the two-point correlator, outlined in \cite{Roberts:2012}, one removes the need for expensive inversions for the tuning. Unfortunately, the optimal level of smearing depends on the quark mass, $\beta$ value, momentum or operator. One must tune the smearing for each set of parameters under consideration. Immediately, one can find appeal in the variational approach as there is no longer a need to tediously tune the operators to match the ground state.

The variational approach provides us with a systematic framework for constructing operators whereby we have not suppressed, but instead removed the contributions of the nearby states. To see how small one could make the variational basis so as to obtain the correct result, we examined all possible subsets of our variational basis. The results are displayed in Fig.~4. To ensure excited state effects are well suppressed it appears that the higher levels of smearing are key. Furthermore, clean results require at least a $3 \times 3$ correlation matrix.

\begin{figure}[t]
	\centering
	\includegraphics{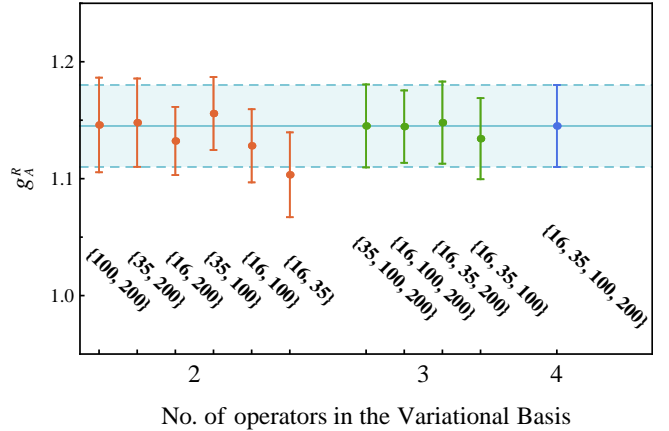}
	\caption{Results for $g_A$ with different number and combinations of operators used in the variational analysis.}
\end{figure}

\section{Cost-Benefit Discussion}

A concern with the correlation matrix approach is the increased cost. For our implementation, we require 2 inversions per configuration for every smearing we include in constructing the correlation matrix. For $n$ smearings we have a total of $2n = 8$ inversions per configuration, as opposed to the minimum of 2. In Fig.~2 we can see that, for large Euclidean times, the conventional approach is consistent with the correlation matrix approach, albeit with larger errors. Thus it is worth considering what the required increase in statistical sample would be for the conventional approach to produce results with similar error to that of our correlation matrix method.

Given the error varies with the sample size $N$ as $\Delta g_A \propto \frac{1}{\sqrt{N}}$ then the relative increase in sample required to obtain an error $(\Delta g_A)_{\textrm{desired}}$ is given by
\[
\frac{N_{\textrm{required}}}{N_{\textrm{current}}} = \left( \frac{(\Delta g_A)_{\textrm{current}}}{(\Delta g_A)_{\textrm{desired}}} \right)^2 = \left( \frac{(\Delta g_A)_{\textrm{sm-sm}}}{(\Delta g_A)_{\textrm{CM}}} \right)^2 \, ,
\]
where $(\Delta g_A)_{\textrm{current}}$ is the error extracted with the current sample of size $N_{\textrm{current}}$. Using the leading time-slice of the associated fit-windows as indicative of the uncertainty in $g_A$, which for the \textit{smeared}-\textit{smeared} approach is $t_s=27$ and for the correlation matrix approach $t_s=23$, we find that
\[
\left.
	\begin{aligned}
	(\Delta g_A)_{\textrm{sm-sm}}&= 0.059 \\
	(\Delta g_A)_{\textrm{CM}}&= 0.030
	\end{aligned}
\quad\right\}
\; \frac{N_{\textrm{required}}}{N_{\textrm{current}}} = \left(\frac{0.59}{0.30}\right)^2 = 3.87 \, .
\]
Naively we expect a factor 4 increase in statistics, which would require fewer inversions than our correlation matrix method. However, we note that the peak value for the \textit{smeared}-\textit{smeared} approach is at time slice 26 and so $\chi^2_{\textrm{dof}}$ analysis would tend to favour earlier points around times 24-25. This is consistent with Table.~2. In the tradition of choosing the earliest possible fit-window to minimise statistical uncertainty, a more appropriate fit window would be times 24-31. Being conscientious of the rapid growth in error bars, the best choice would be times 25-28 with $\chi^2_{\textrm{dof}} =$ 0.9 and a result $g_A =$ 1.38(3). This result is systematically suppressed, relative to the correct result of $g_A =$ 1.47(2), by excited state effects. While one could invest more super-computing resources to reduce statistical error, in this case one will only get the wrong answer very accurately if one does not take care in fine-tuning the source.

To further understand this we note that using the variational approach, ground state domination occurs earlier in Euclidean time thus allowing the current insertion at an earlier time. For this particular ensemble, ground state dominance for the nucleon occurs at time $t=21$, so our choice for $t_C$ is ideal for the correlation matrix method. For the \textit{smeared}-\textit{smeared} approach with 35 sweeps of smearing, ground state dominance does not occur until time $t=23$. This is why the peak value is systematically low. The downwards shift for small source smearings is the result of the current being too early, sampling both ground state and excited state contributions to the matrix element. This also gives rise to the smearing dependence illustrated in Fig.~3. Thus for a more comprehensive comparison, one requires a new simulation with $t_C=23$, two time slices later. However, we can still get some insight from our present analysis into the required increase in statistics. For the the ratio of three- to two-point functions, ground state dominance occurs 6 time slices after the current insertion, so with $t_C=23$ one would be considering a fit window commencing at $t_s=29$ as opposed to $t_s=27$ considered earlier. Here we have
\[
\hspace{-8pt}
\left.
	\begin{aligned}
	(\Delta g_A)_{\textrm{sm-sm}}&= 0.101 \\
	(\Delta g_A)_{\textrm{CM}}&= 0.030
	\end{aligned}
\quad\right\}
\; \frac{N_{\textrm{required}}}{N_{\textrm{current}}} = \left(\frac{0.101}{0.030}\right)^2 \simeq 11.3 ,
\]
a factor 11 increase.

As the variational approach enables one to:
\begin{enumerate}
	\item rapidly isolate the ground state following the source, thus enabling an earlier current insertion, and
	\item rapidly isolate the ground state again after inserting the current enabling an earlier Euclidean time fit,
\end{enumerate}
the associated reduction in the error bar through this process outweighs the increased cost in constructing the matrix of cross-correlators.

In our implementation, due to the construction of the complete correlation matrix of three-point functions, we not only have access to the ground state, but also to the first $n-1$ excited states, where $n$ is the dimension of our operator basis. This has been utilised in \cite{Maurer:2012} to access the axial charge of nucleon excitations. In principle, if one were solely interested in the ground state properties, one could use the optimised sources generated via the two-point correlation matrix as the input for the SST inversion, providing SST propagators that couple directly with the ground state. This reduces the cost from $2n$ inversions down to $n+1$. For this calculation the cost would be reduced from 8 to 5 inversions. Further reduction in cost is demonstrated through Fig.~4. It was found that access to ground state properties can be achieved with 3 levels of smearing, provided the smearings are chosen to span the space. Therefore, we could further reduce the cost to $3+1=4$ inversions per configuration, only a factor of 2 above the minimum for what is equivalent to an order of magnitude improvement in the statistics.

In Table 3 we present a comparison of our result for $g_A^R$ with results by other groups on similar ensembles. The consistency between our result and those of other groups is testament to the care taken by these collaborations to minimise systematic uncertainties.

\begin{table}[h!]
	\centering
	\caption{Comparison of results for $g_A^R$ on ensembles with similar volumes and values of $m_\pi$ to our calculation. For the CLS/Mainz group we have included results for the conventional ratio method (upper) and the summation method (lower). The asterisk indicates that these results include the correction of finite-volume effects and so will tend to sit slightly higher. }
	\hspace*{-6pt}
	\scalebox{0.9}{
	\begin{tabular}{lccccc}
	\hline\noalign{\smallskip}
	Group & $m_{\pi}$ (MeV) & $m_{\pi}L$ & $t_{s} - t_0$ (fm) & $g_A^R$ &\\
	\hline\noalign{\smallskip}
	our result & 290 & 4.26 & 0.75 & 1.147(33) &\\
	\noalign{\smallskip}
	QCDSF '13     & 292 & 4.25 & 1.1 & 1.099(13) & \\
	CLS/Mainz '12 & 277 & 4.25 & 1.1 & 1.137(37) & *\\
	CLS/Mainz '12 & 277 & 4.25 & 0.7-1.3 & 1.162(95) & *\\
	LHPC '10      & 293 & 3.68 & 1.2 & 1.154(26) &\\
	ETMC '10      & 298 & 4.28 & 1.1 & 1.103(32) &\\
	\noalign{\smallskip}
	\hline
	\end{tabular}}
\end{table}

A key issue in the calculation of any three-point function is how large must one make their source-sink separation to ensure that excited state contaminations are sufficiently suppressed \cite{Dinter:2011}. There is a general consensus within the community that source-sink separations $\lesssim 1.0$ fm will suffer from excited state contaminations without fine-tuning the source and sink to isolate the state. Indeed our results highlight this systematic effect when using the conventional approach. Here the source-sink separation of $\sim 1.0$ fm is too small and the extracted value for $g_A$ suffers from excited state effects as illustrated in Fig.~3. The underlying issue is that there is insufficient time to isolate the ground state prior to current insertion and again isolate the ground state before annihilation. Based on our earlier arguments regarding a more suitable current insertion time, we would expect an suitable sink time would be $t_{s} = 29$, increasing the source-sink separation to $\sim 1.2$ fm. This result is consistent with the source-sink separations used by the other groups in Table 3.

Using the variational approach, due to rapid onset of ground state dominance through ideal interpolators, we are able to use much smaller source-sink separations. For our variational results, ground state dominance after the current insertion occurs as early as $t_{s} = 23$ resulting in a temporal separation between source and sink of only 0.64 fm. Thus, by applying the variational technique to fixed sink methods, one could consider source-sink separations $\sim 0.7$ fm which would result in small statistical errors.

\section{Conclusion}

In this letter we have illustrated how the variational approach can be used to eliminate excited state effects from the calculation of the nucleon axial-vector coupling constant $g_A$. These effects act to suppress lattice simulation results for $g_A$. The use of optimised interpolators results in rapid ground state dominance allowing for earlier insertion of the current and earlier fit windows resulting in smaller statistical uncertainty. The key advantage to this approach is that once a suitable basis has been chosen, optimised sources are constructed automatically eliminating the need to tune smearing parameters and source-sink separations.

The method is general and would be ideally suited to calculations of form factors where the variational approach could be applied separately for each choice of source-sink momentum combination. Another quantity that has so far proved elusive for lattice calculations and could benefit from our approach is the quark momentum fraction, $\langle x \rangle$, which is notorious for producing lattice results that are more than $50\%$ larger than phenomenological determinations (see \cite{Renner:2011} for a review).

Future investigations will accurately calculate $g_A$ at a variety of quark masses and connect these results to Nature via finite-volume chiral effective field theory.

\section*{Acknowledgements}
\sloppy We thank the PACS-CS Collaboration for making available the $2+1$ flavor configurations used in this analysis and the ILDG for creating the framework to access this data. This research was undertaken with the assistance of resources at the NCI National Facility in Canberra, Australia, and the iVEC facilities at Murdoch University (iVEC@Murdoch) and the University of Western Australia (iVEC@UWA). These resources were provided through the National Computational Merit Allocation Scheme, supported by the Australian Government. This research is supported by the Australian Research Council.



\bibliographystyle{model1-num-names}
\bibliography{./gA}

\end{document}